\documentclass[11pt,a4paper]{article}
\pdfoutput=1
\usepackage{jheppub}
\usepackage{tikz}
\usepackage{subcaption}
\DeclareMathOperator{\Tr}{Tr}

\newcommand{\hf}{\frac{1}{2}}

\newcommand{\si}{\sigma}

\newcommand{\del}{\partial}

\newcommand{\bra}{\langle}
\newcommand{\ket}{\rangle}

\newcommand{\bt}{\beta}

\newcommand{\rt}[1]{\sqrt{#1}}
\newcommand{\cO}{\mathcal{O}}
\newcommand{\cZ}{\mathcal{Z}}

\begin{document}

\title{Quenched free energy in random matrix model}

\author{Kazumi Okuyama}

\affiliation{Department of Physics, Shinshu University,\\
3-1-1 Asahi, Matsumoto 390-8621, Japan}

\emailAdd{kazumi@azusa.shinshu-u.ac.jp}

\abstract{
We compute the quenched free energy in the Gaussian random matrix model
by directly evaluating the matrix integral without using the replica trick.
We find that the quenched free energy
is a monotonic function of the temperature and the entropy
approaches $\log N$ at high temperature and vanishes at zero temperature.
}

\maketitle

\section{Introduction}\label{sec:intro}
Recently, it is shown that the path integral
of Jackiw-Teitelboim (JT) gravity \cite{Jackiw:1984je,Teitelboim:1983ux}
is equivalent to a certain double-scaled matrix model \cite{Saad:2019lba}.
From the viewpoint of holography, this implies that
the holographic dual of JT gravity is not a single quantum mechanical system but
an ensemble of systems with random Hamiltonians. 
Moreover, it is realized in \cite{Saad:2018bqo}  
that the Euclidean wormhole connecting different boundaries of spacetime
plays an important role in explaining the so-called ``ramp''
of the spectral form factor \cite{Garcia-Garcia:2016mno,Cotler:2016fpe}
in the Sachdev-Ye-Kitaev (SYK) model \cite{Sachdev,kitaev2015simple}.
The importance of the wormhole
in quantum gravity is also emphasized in the recent computation of the Page
cure using the replica method \cite{Penington:2019kki,Almheiri:2019qdq},
where the inclusion of the so-called replica wormhole is essential
for the resolution of the apparent paradox in the original Hawking's calculation.\footnote{However, the recovering of a unitary Page curve is far from sufficient to completely resolve the paradox. We still lack an understanding of by what mechanism information is able to exit an evaporating black hole.} 

In recent papers \cite{Engelhardt:2020qpv,Johnson:2020mwi}
the replica method is applied to
the computation of the free energy in JT gravity.
As emphasized in \cite{Engelhardt:2020qpv}, this problem is very interesting to reveal
the role of replica wormholes and explore the possibility of 
replica symmetry breaking 
and the putative 
spin glass phase of quantum gravity.\footnote{The replica symmetry breaking in the SYK model is discussed in 
\cite{Gur-Ari:2018okm,Arefeva:2018vfp}.
The idea of replica symmetry breaking is developed by Parisi
\cite{parisi1979infinite,parisi1980sequence} to solve the sping glass model
of Sherrington and Kirkpatrick \cite{sherrington1975solvable}.
See e.g. \cite{Denef:2011ee,castellani2005spin,sherrington} for reviews of spin glasses.}
However, it is reported in \cite{Engelhardt:2020qpv} that a naive application
of the replica method leads to a pathological behavior of the free energy.
In a recent paper \cite{Johnson:2020mwi} it
is emphasized that the non-perturbative effect is important to resolve this
problem.

In this paper, we will consider a simple toy model for
the computation of free energy in JT gravity.
Instead of the matrix model of JT gravity in \cite{Saad:2019lba},
we consider the free energy in the Gaussian matrix model where
the Hamiltonian is regarded as a random hermitian matrix
with Gaussian distribution.\footnote{This problem is suggested
in the discussion section in \cite{Engelhardt:2020qpv}.}
We are interested in the the so-called quenched free energy 
$\bra \log Z(\bt)\ket$\footnote{
Strictly speaking the quenched free energy is defined by including the factor of temperature $F=-T\bra \log Z(\bt)\ket$ as in \eqref{eq:def-FS}, but we will loosely use the name 
``quenched free energy'' to indicate either $\bra \log Z(\bt)\ket$ or $F=-T\bra \log Z(\bt)\ket$ depending on the context. We believe that which one we are referring to is clear from the context and this will not cause a confusion to the readers.
} in the matrix model, where $Z(\bt)=\Tr e^{-\bt H}$
is the partition function with the inverse temperature $\bt=T^{-1}$
and the expectation value is defined by the integral over the $N\times N$
hermitian matrix $H$
\begin{equation}
\begin{aligned}
 \bra f(H)\ket=
\frac{\int dH e^{-\frac{N}{2}\Tr H^2}f(H)}{\int dH e^{-\frac{N}{2}\Tr H^2}}.
\end{aligned} 
\end{equation} 
One can compute the quenched free energy by the replica method\footnote{
The correlators of the resolvent $\Tr(E-H)^{-1}$ in the Gaussian matrix
model are analyzed by the replica method in \cite{kamenev1999wigner}.
It is found that the replica symmetry breaking is important to reproduce
the known results of the correlators of resolvents.
In this paper we are dealing with the different quantity
$\bra\log Z(\bt)\ket$ and the computation in \cite{kamenev1999wigner}
cannot simply be generalized to our case.
}
\begin{equation}
\begin{aligned}
 \bra \log Z(\bt)\ket=\lim_{n\to0}\frac{\bra Z(\bt)^n\ket-1}{n}.
\end{aligned} 
\label{eq:replica}
\end{equation}
In the high temperature regime the $n$-point correlator
$\bra Z(\bt)^n\ket$ is approximated by the disconnected correlator
\begin{equation}
\begin{aligned}
 \bra Z(\bt)^n\ket\approx\bra Z(\bt)\ket^n,
\end{aligned} 
\end{equation}
and the $n\to0$ limit in \eqref{eq:replica} gives rise to
\begin{equation}
\begin{aligned}
 \bra \log Z(\bt)\ket\approx\lim_{n\to0}\frac{\bra Z(\bt)\ket^n-1}{n}
=\log\bra Z(\bt)\ket.
\end{aligned} 
\label{eq:ann}
\end{equation}
The right hand side of this equation is known as the annealed free energy.
On the other hand, in the low temperature regime it is not clear how to
define the analytic continuation of
$\bra Z(\bt)^n\ket$ to $n<1$.
This is the origin of the difficulty found in \cite{Engelhardt:2020qpv}.

It turns out that we can avoid this difficulty of analytic continuation
by directly evaluating the quenched free energy by
the matrix integral
\begin{equation}
\begin{aligned}
 \bra \log Z(\bt)\ket=\frac{\int dH e^{-\frac{N}{2}\Tr H^2}\log\Tr e^{-\bt H}}{\int dH e^{-\frac{N}{2}\Tr H^2}}.
\end{aligned} 
\label{eq:quench-GUE}
\end{equation}
We can rewrite this integral \eqref{eq:quench-GUE}
as an integral over the $N$ eigenvalues of the matrix $H$
and study the physical quantities like
the free energy $F$ and the entropy $S$
\begin{equation}
\begin{aligned}
 F=-T\bra \log Z(\bt)\ket,\quad S=-\frac{\del F}{\del T}.
\end{aligned} 
\label{eq:def-FS}
\end{equation}
In order for the entropy to be positive, the free energy $F$ should be a monotonically
decreasing function of $T$.
In the replica computation of the quenched free energy of JT
gravity \cite{Engelhardt:2020qpv}, a pathological non-monotonic behavior
of $F$ is found under a certain prescription of the analytic
continuation in $n$.\footnote{As emphasized in \cite{Engelhardt:2020qpv} the analytic continuation of
$\bra Z(\bt)^n\ket$ from positive integer $n$ to $n<1$ is not unique.
The non-monotonic behavior of the free energy is a consequence of the particular choice of the analytic continuation used in \cite{Engelhardt:2020qpv}.
However, as explained in \cite{Engelhardt:2020qpv}
their choice of analytic continuation 
is not meant to be the correct one but it is just an illustrative example
to demonstrate the importance of
the replica wormholes in the computation of free energy.}
We find that the direct computation of the quenched
free energy in the Gaussian matrix model \eqref{eq:quench-GUE}
gives rise to a well-defined monotonic behavior of the free energy $F$.

This paper is organized as follows. In section \ref{sec:quench},
we find the explicit integral representation of the quenched free energy
\eqref{eq:quench-GUE} and study its behavior in the high and low temperature regimes.
In section \ref{sec:num}, we study the exact free energy and entropy
for $N=2,3$ as examples.
We find that the free energy exhibits a well-defined monotonic behavior 
as a function of $T$.
In section \ref{sec:replica}, we comment on the computation using the replica method.
We propose a necessary condition for the
analytic continuation of $\bra Z(\bt)^n\ket$ to satisfy. 
Finally, we conclude in section \ref{sec:discussion} with some discussions 
on the interesting future problems.
\section{Quenched free energy in Gaussian matrix model}\label{sec:quench}
In this paper we will analyze the quenched free energy in Gaussian 
matrix model \eqref{eq:quench-GUE} directly without using the replica trick.
From the standard argument, the matrix integral in \eqref{eq:quench-GUE}
is written as an integral over the $N$ eigenvalues $\{E_1,\cdots, E_N\}$ of $H$
\begin{equation}
\begin{aligned}
 \bra \log Z(\bt)\ket=\frac{1}{\cZ}\frac{1}{N!}\int_{-\infty}^\infty\prod_{i=1}^N
dE_ie^{-\frac{N}{2}E_i^2}\prod_{i<j}(E_i-E_j)^2\log\left(
\sum_i e^{-\bt E_i}\right).
\end{aligned} 
\label{eq:quench-F}
\end{equation}
Here the normalization factor $\cZ$ is given by
\begin{equation}
\begin{aligned}
 \cZ=\frac{1}{N!}\int_{-\infty}^\infty\prod_{i=1}^N
dE_ie^{-\frac{N}{2}E_i^2}\prod_{i<j}(E_i-E_j)^2=N^{-\frac{N^2}{2}}
(2\pi)^{\frac{N}{2}}G_2(N+1),
\end{aligned} 
\end{equation}
where $G_2(N+1)$ denotes the Barnes $G$-function.
Using this expression \eqref{eq:quench-F}, in subsection \ref{sec:high} and \ref{sec:low} we will study the behavior of
quenched free energy in the high temperature and the low temperature regimes,
respectively.

\subsection{High temperature regime}\label{sec:high}
In the high temperature regime, the quenched free energy is approximated 
by  the annealed free energy \eqref{eq:ann}.

The one-point function $\bra Z(\bt)\ket$ in the Gaussian
matrix model 
happens to be the same as the computation of the $1/2$ BPS Wilson loop 
in $\mathcal{N}=4$ super Yang-Mills theory (SYM), and the exact result at finite
$N$ is found in \cite{Drukker:2000rr}
in terms of the Laguerre polynomial
\begin{equation}
\begin{aligned}
 \bra Z(\bt)\ket=e^{\frac{\bt^2}{2N}}L^1_{N-1}\left(-\frac{\bt^2}{N}\right).
\end{aligned} 
\label{eq:Z-Lag}
\end{equation}
The large $N$ behavior of the one-point function $\bra Z(\bt)\ket$ 
can be computed from the genus-zero eigenvalue density 
\begin{equation}
\begin{aligned}
 \rho_0(E)=\frac{1}{2\pi}\rt{4-E^2},
\end{aligned} 
\label{eq:wigner}
\end{equation}
known as the Wigner semi-circle distribution. Then in the large $N$ limit
the one-point function $\bra Z(\bt)\ket$ becomes
\begin{equation}
\begin{aligned}
 \bra Z(\bt)\ket&\approx N\int_{-2}^2dE\rho_0(E)e^{-\bt E}=N\frac{I_1(2\bt)}{\bt},
\end{aligned}
\label{eq:Z-g0} 
\end{equation}
where $I_1(2\bt)$ is the modified Bessel function of the first kind.
From this expression one can easily find the expansion of 
the free energy and the entropy in the high temperature regime $(T\gg1)$
\begin{equation}
\begin{aligned}
F&=-T\log N-\hf T^{-1}+\cO(T^{-3}),\\
S&=\log N-\hf T^{-2}+\cO(T^{-4}).
\end{aligned} 
\end{equation}
In particular, the high temperature limit of entropy is $\log N$
\begin{equation}
\begin{aligned}
 \lim_{T\to\infty}S=\log N.
\end{aligned} 
\end{equation}
This is expected since $N$ is the dimension of the Hilbert space
and $\log N$ is the maximal entropy of the system.

\subsection{Low temperature regime}\label{sec:low}
Next let us consider the low temperature regime $(T\ll1)$.
In the low temperature limit $\bt\to\infty$, one can see that
$\log\Tr e^{-\bt H}$ becomes
\begin{equation}
\begin{aligned}
 \lim_{\bt\to\infty}\log\left(\sum_{i=1}^Ne^{-\bt E_i}\right)=-\bt\text{min}\{E_i\},
\end{aligned} 
\end{equation}
where $\text{min}\{E_i\}$ denotes the smallest eigenvalue
in the set of $N$ eigenvalues $\{E_1,\cdots, E_N\}$.
Thus we find that the low temperature limit of the
quenched free energy is determined by the
expectation value $E_0=\bra \text{min}\{E_i\}\ket$ of the smallest eigenvalue
\begin{equation}
\begin{aligned}
 \lim_{\bt\to\infty}\bra\log Z(\bt)\ket=-\bt E_0.
\end{aligned} 
\label{eq:lowT-lim}
\end{equation}
Note that $E_0$ is explicitly written as the eigenvalue integral
\begin{equation}
\begin{aligned}
 E_0=\frac{1}{\cZ}\frac{1}{N!}
\int_{-\infty}^\infty\prod_{i=1}^N
dE_ie^{-\frac{N}{2}E_i^2}\prod_{i<j}(E_i-E_j)^2\text{min}\{E_i\}.
\end{aligned} 
\end{equation}
We do not know the closed form of this integral for general $N$,
but it is possible to evaluate this integral for small $N$. For instance, 
for $N=2,3$ we find 
\begin{equation}
\begin{aligned}
 E_0^{(N=2)}&=-\rt{\frac{2}{\pi}}\approx -0.79788,\\
E_0^{(N=3)}&=-\frac{9\rt{3}}{8\rt{\pi}}\approx -1.09936.
\end{aligned} 
\end{equation}
It is known \cite{bai1988necessary}
that 
in the large $N$ limit $E_0$ 
converges to the edge of the Wigner semi-circle 
distribution \eqref{eq:wigner}
\begin{equation}
\begin{aligned}
 E_0^{(N=\infty)}=-2.
\end{aligned} 
\label{eq:E0-lim}
\end{equation}

It turns out that one can systematically compute the
small $T$ corrections to the leading term in \eqref{eq:lowT-lim}.
In the eigenvalue integral \eqref{eq:quench-F}, one can choose
$E_N$ as the smallest eigenvalue without loss of generality. Then 
the range of other eigenvalues $E_i~(i=1,\cdots, N-1)$ is restricted to $E_i>E_N$.
With this remark in mind, the quenched free energy is written as
\begin{equation}
\begin{aligned}
 \bra\log Z(\bt)\ket&=-\bt E_0+\frac{1}{\cZ}\frac{N}{N!}\int_{-\infty}^\infty dE_N
e^{-\frac{N}{2}E_N^2}\int_{E_N}^\infty\prod_{i=1}^{N-1}
dE_ie^{-\frac{N}{2}E_i^2}\\
&\times \prod_{i=1}^{N-1}(E_i-E_N)^2\prod_{1\leq i<j\leq N-1}(E_i-E_j)^2 
\log\left(1+\sum_{i=1}^{N-1}e^{-\bt(E_i-E_N)}\right).
\end{aligned} 
\end{equation} 
This is further simplified by shifting $E_i\to E_i+E_N~(i=1,\cdots,N-1)$
and integrating out $E_N$ 
\begin{equation}
\begin{aligned}
 \bra\log Z(\bt)\ket&=-\bt E_0+\frac{1}{\cZ}
\frac{\rt{2\pi}}{N!}
\int_{0}^\infty\prod_{i=1}^{N-1}
dE_ie^{-\frac{N}{2}\sum_iE_i^2+\hf(\sum_iE_i)^2}\\
&\times \prod_{i=1}^{N-1}E_i^2\prod_{1\leq i<j\leq N-1}(E_i-E_j)^2 
\log\left(1+\sum_{i=1}^{N-1}e^{-\bt E_i}\right).
\end{aligned} 
\label{eq:F-int}
\end{equation}
This is our master formula.

The small $T$ behavior of \eqref{eq:F-int}
is found by rescaling one of the integration variables $E_i\to TE_i$.
In this way we find that the quenched free energy at low temperature 
$(T\ll1)$ behaves as
\begin{equation}
\begin{aligned}
 F=-T\bra \log Z(\bt)\ket=E_0-\si T^{4}+\cO(T^{5}),
\end{aligned} 
\label{eq:F-lowN}
\end{equation}
where $\si$ is an $N$-dependent constant.
From \eqref{eq:F-lowN}, it follows that the entropy in the low temperature regime behaves as
\begin{equation}
\begin{aligned}
 S=-\frac{\del F}{\del T}=4\si T^{3}+\cO(T^{4}).
\end{aligned} 
\end{equation}
This result implies that the entropy vanishes at zero temperature
\begin{equation}
\begin{aligned}
 \lim_{T\to0}S=0.
\end{aligned} 
\label{eq:S-zero}
\end{equation}
Note that the vanishing
of the entropy at zero temperature is also observed in
the Parisi's solution of 
Sherrington-Kirkpatrick model \cite{crisanti2002analysis,sommers1984distribution}.
\section{\mathversion{bold}Numerics for $N=2,3$}\label{sec:num}
\begin{figure}[htb]
\centering
\subcaptionbox{Free energy at $N=2$ \label{sfig:F2}}{\includegraphics[width=0.45\linewidth]{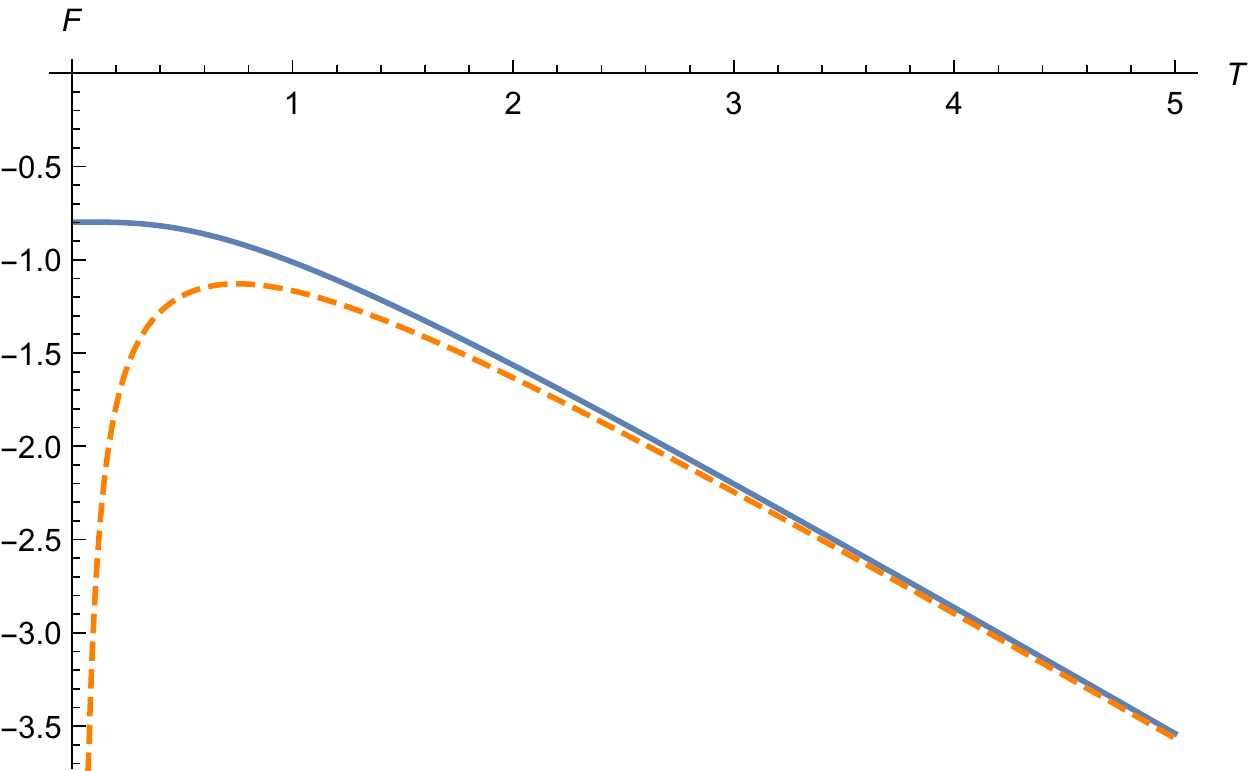}}
\hskip5mm
\subcaptionbox{Free energy at $N=3$ \label{sfig:F3}}{\includegraphics[width=0.45\linewidth]{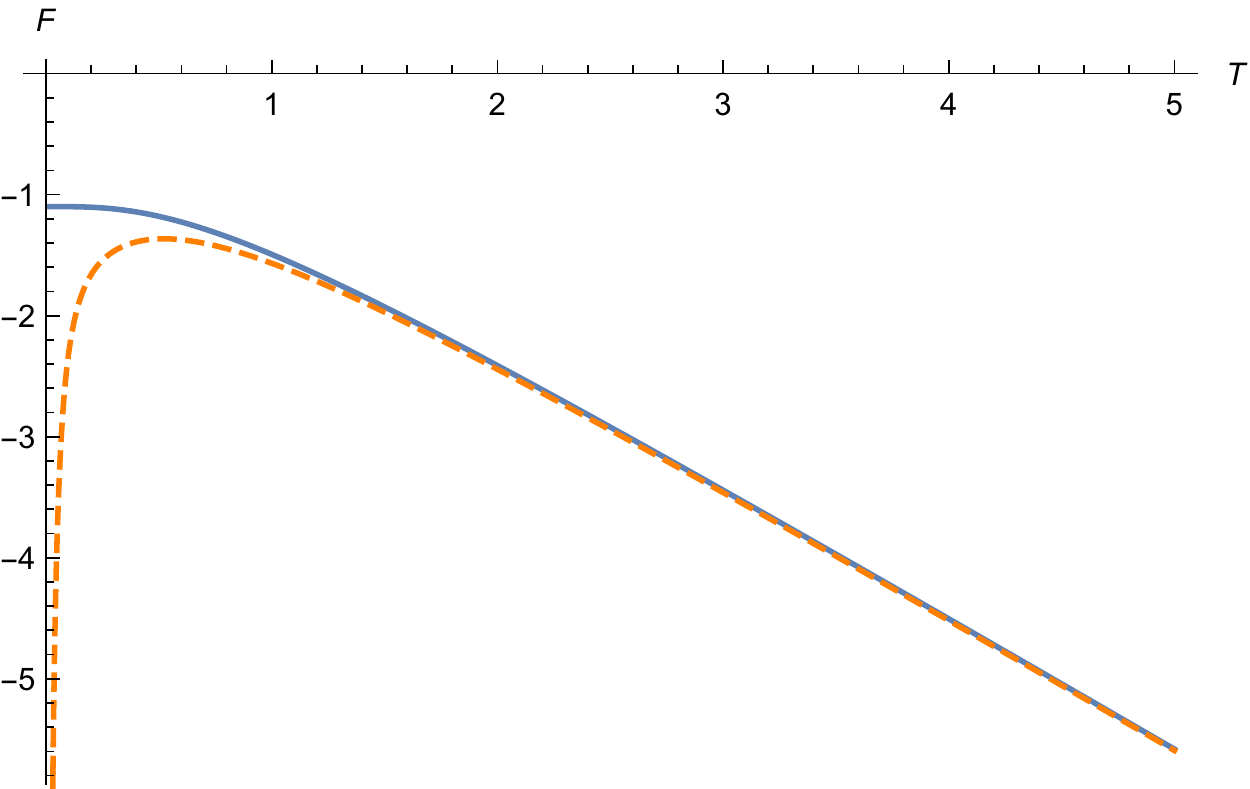}}
  \caption{
Plot 
of free energy for \subref{sfig:F2} $N=2$ and \subref{sfig:F3} $N=3$
as a function of temperature $T$.
The solid curves are the quenched free energy while the orange dashed curves represent
the annealed free energy $F_{\text{ann}}=-T\log \bra Z(\bt)\ket$ with
the exact one-point function in \eqref{eq:Z-Lag}.
}
  \label{fig:F23}
\end{figure}

\begin{figure}[htb]
\centering
\subcaptionbox{Entropy at $N=2$ \label{sfig:S2}}{\includegraphics[width=0.45\linewidth]{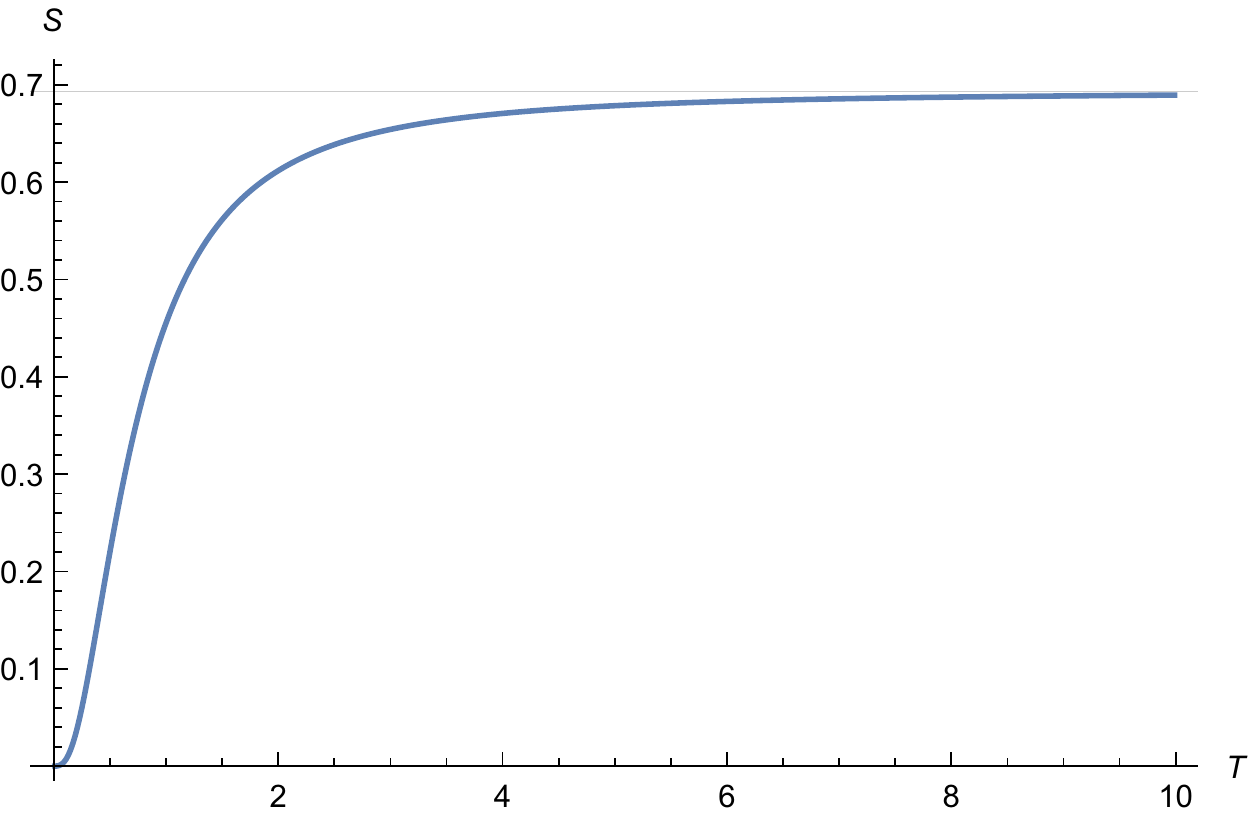}}
\hskip5mm
\subcaptionbox{Entropy at $N=3$ \label{sfig:S3}}{\includegraphics[width=0.45\linewidth]{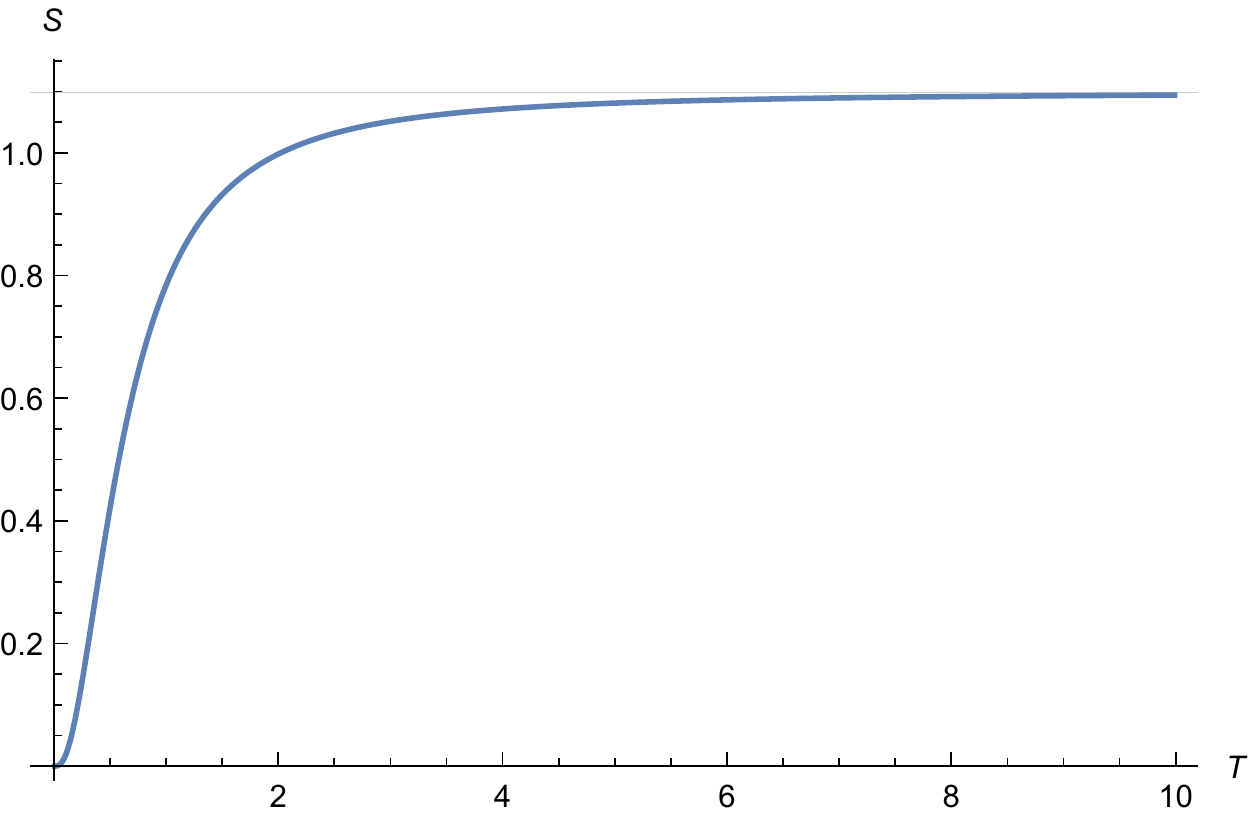}}
  \caption{
Plot 
of entropy for \subref{sfig:S2} $N=2$ and \subref{sfig:S3} $N=3$
as a function of temperature $T$.
The solid curves are the exact result while the horizontal gray lines 
represent the maximum value of entropy $S=\log N$.
}
  \label{fig:S23}
\end{figure}

In this section we study numerically the 
integral representation of the quenched free energy \eqref{eq:F-int}
for $N=2,3$.
For $N=2$ the integral \eqref{eq:F-int} becomes 
\begin{equation}
\begin{aligned}
 \bra \log Z(\bt)\ket=\rt{\frac{2}{\pi}}\bt
+\rt{\frac{2}{\pi}}\int_0^\infty dE e^{-\hf E^2}E^2\log(1+e^{-\bt E}),
\end{aligned} 
\label{eq:exact-N2}
\end{equation}
and for $N=3$ we find
\begin{equation}
\begin{aligned}
  &\bra \log Z(\bt)\ket=\frac{9\rt{3}}{8{\rt{\pi}}}\bt\\
+&\frac{27\rt{3}}{8\pi}\int_0^\infty dE_1
\int_0^\infty dE_2 e^{-\frac{3}{2}(E_1^2+E_2^2)+\hf(E_1+E_2)^2}E_1^2E_2^2(E_1-E_2)^2\log(1+e^{-\bt E_1}+e^{-\bt E_2}).
\end{aligned} 
\end{equation}
One can easily evaluate these integrals numerically.
In Fig.~\ref{fig:F23} we show the plot of free energy as a function of temperature.
At high temperature, the quenched free energy approaches the
annealed free energy $F_{\text{ann}}=-T\log\bra Z(\bt)\ket$ (orange dashed curve)
as expected.
In Fig.~\ref{fig:S23} we show the plot of entropy $S$.
One can see that $S$ approaches $\log N$ at high temperature and vanishes at
zero temperature.

\begin{figure}[htb]
\centering
\includegraphics[width=0.8\linewidth]{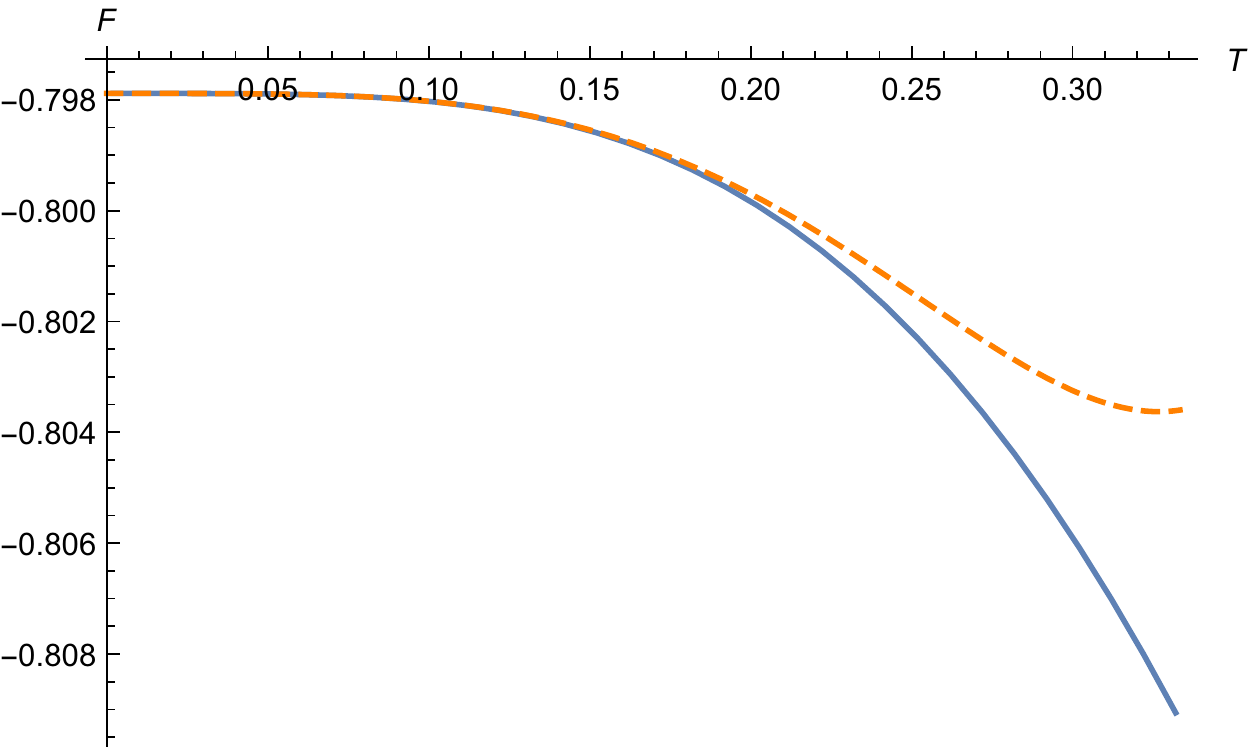}
  \caption{
Plot 
of free energy for $N=2$ at low temperature.
The solid curve is the exact result while the orange dashed curve 
represents the small $T$ expansion \eqref{eq:F2-low}.
}
  \label{fig:F2-low}
\end{figure}

Let us take a closer look at the low temperature regime 
for $N=2$. The small $T$ expansion of the integral \eqref{eq:exact-N2} is obtained
by
rescaling the integration variable $E\to TE$ and expanding the Gaussian factor in 
\eqref{eq:exact-N2}
\begin{equation}
\begin{aligned}
 F&=-\rt{\frac{2}{\pi}}\Biggl[1+T^4\int_0^\infty dE E^2\sum_{n=0}^\infty\frac{1}{n!}
\left(-\frac{T^2E^2}{2}\right)^n\log(1+e^{-E})\Biggr]\\
&=-\rt{\frac{2}{\pi}}\Biggl[1+\frac{7\pi^4}{360}T^4-\frac{31\pi^6}{2520}T^6
+\cO(T^8)\Biggr].
\end{aligned} 
\label{eq:F2-low}
\end{equation}
Note that the first small $T$ correction is of order $T^4$ which is consistent with
the general result in \eqref{eq:F-lowN}.
In Fig.~\ref{fig:F2-low} we show the plot of quenched free energy for $N=2$
and its small $T$ expansion up to $T^6$ in \eqref{eq:F2-low}.
One can see that the exact quenched free energy is a monotonic function of $T$
even in the low temperature regime and $F$ becomes $E_0$ at zero temperature.
A pathological non-monotonic behavior found in 
\cite{Engelhardt:2020qpv} using the replica trick
does not occur in the exact result of quenched free energy.

\section{Comment on the replica method}\label{sec:replica}
Let us compare our direct calculation of quenched
free energy with the replica method \eqref{eq:replica}.
As we mentioned in section \ref{sec:intro}, one can easily apply the replica method
in the high temperature regime and obtain the 
result \eqref{eq:ann}.
In particular,
in the high temperature limit $\bt\to0$, the partition function 
$Z(\bt)=\Tr e^{-\bt H}$ reduces to the dimension of the Hilbert space
\begin{equation}
\begin{aligned}
 \lim_{\bt\to0}Z(\bt)=\Tr 1=N.
\end{aligned} 
\end{equation}
Thus the quenched free energy approaches the maximal entropy of the system
in the limit $T\to\infty$
\begin{equation}
\begin{aligned}
 \lim_{\bt\to0}\bra \log Z(\bt)\ket=\lim_{n\to0}\frac{N^n-1}{n}=\log N.
\end{aligned} 
\end{equation}

On the other hand, the application of the replica trick in the low temperature regime
is rather subtle. Under a certain prescription of the analytic continuation in 
the number of replicas $n$, it is found that 
the free energy exhibits a non-monotonic behavior 
as a function of temperature \cite{Engelhardt:2020qpv}.

Our direct computation of the quenched free energy puts a certain constraint
on the possible form of the analytic continuation in $n$.
At low temperature, the smallest eigenvalue $E_0$ of $H$ becomes dominant
and thus we expect
\begin{equation}
\begin{aligned}
 \lim_{\bt\to\infty}\bra Z(\bt)^n\ket= e^{-n\bt E_0}.
\end{aligned} 
\label{eq:condition}
\end{equation}
We can regard \eqref{eq:condition} as a condition for the possible analytic continuation
of $\bra Z(\bt)^n\ket$ to satisfy.
Then we can apply the replica method in the low temperature regime
\begin{equation}
\begin{aligned}
 \lim_{\bt\to\infty}\bra \log Z(\bt)\ket=\lim_{n\to0}\frac{e^{-n\bt E_0}-1}{n}=-\bt E_0,
\end{aligned} 
\label{eq:low-replica}
\end{equation}
which reproduces the correct behavior of the quenched free energy \eqref{eq:lowT-lim}.

Note that there is no $\log N$ entropy term in \eqref{eq:low-replica}
since only a single eigenvalue 
(the lowest energy state) contributes to $\bra Z(\bt)^n\ket$ 
in the low temperature limit. 
This explains the vanishing of entropy at zero temperature \eqref{eq:S-zero}.

We would like to understand the role of replica symmetry breaking
in a possible large $N$ phase transition. 
When $n$ is a positive integer, the $n$-replica correlator $\bra Z(\bt)^n\ket$
is expanded in terms of the connected correlators 
\begin{equation}
\begin{aligned}
\bra Z(\bt)^n\ket=
n!\sum_{\sum p\nu_p=n}\prod_{p=1}^n \frac{1}{\nu_p!(p!)^{\nu_p}}
\Bigl(\bra Z(\bt)^p\ket_{\text{conn}}\Bigr)^{\nu_p}.
\end{aligned} 
\label{eq:Zin-Zconn}
\end{equation}
Here $[p^{\nu_p}]=[1^{\nu_1}2^{\nu_2}\cdots n^{\nu_n}]$ denotes a partition of $n$.
In the high temperature regime the disconnected part $\bra Z(\bt)\ket^n$
corresponding to the partition $[1^n]$ is dominant, while 
at low temperature the totally connected part
$\bra Z(\bt)^n\ket_{\text{conn}}$ corresponding to the partition $[n^1]$ is dominant
\cite{Okuyama:2019xvg}.
Then one might naively think that the quenched free energy in
the low temperature regime is given by
the totally connected correlator $\bra Z(\bt)^n\ket_{\text{conn}}$
\begin{equation}
\begin{aligned}
 \bra \log Z(\bt)\ket=\lim_{n\to0}\frac{\bra Z(\bt)^n\ket_{\text{conn}}-1}{n}.
\end{aligned} 
\label{eq:lim-conn}
\end{equation}
One can try to compute $\bra Z(\bt)^n\ket_{\text{conn}}$ for integer
$n$ and analytically continue it to $n=0$.
However, this analytic continuation is very subtle since
$\bra Z(\bt)^n\ket_{\text{conn}}$ scales as $N^{2-n}$ in the large $N$ limit
and the naive $n\to0$ limit of $\bra Z(\bt)^n\ket_{\text{conn}}$ is not $1$
and the limit \eqref{eq:lim-conn} does not exist.
It is not clear how to define the analytic continuation of
$\bra Z(\bt)^n\ket_{\text{conn}}$ which satisfies the condition \eqref{eq:condition}.
We believe that \eqref{eq:lim-conn} is \textit{not} the correct way to compute
the low temperature regime of quenched free energy.
In other words, the two limits $\bt\to\infty$ and $n\to0$ do not commute.
 
A similar problem has appeared in the so-called 
random energy model \cite{derrida1981random}.\footnote{
The random energy model is defined as a model with $N$ randomly distributed energy
eigenvalues with Gaussian distribution but the correlation among eigenvalues is
ignored.
It is known that the random energy model is equivalent to the $p\to\infty$
limit of a $p$-spin generalization of the Sherrington-Kirkpatrick model
\cite{derrida1980random,derrida1981random}. 
}
In \cite{derrida1981random} this problem is circumvented by
promoting $(p,\nu_p)$ in \eqref{eq:Zin-Zconn} as a continuous variable
and the correct low temperature behavior is obtained by plugging $\nu_p=\frac{n}{p}$ and
extremizing the term $(\bra Z(\bt)^p\ket_{\text{conn}})^{n/p}$ 
in \eqref{eq:Zin-Zconn} with respect to $p$.
The $n$-point function $\bra Z(\bt)^n\ket$ obtained with this prescription
indeed satisfies the necessary condition \eqref{eq:condition} 
and we can safely take the $n\to0$ limit \cite{derrida1981random}.
The resulting quenched free energy $F$ exhibits a phase transition  in the large $N$
limit at a certain critical temperature $T_c$: for $T>T_c$, $F$ agrees with the
annealed free energy which takes the form $F_{\text{ann}}=aT+bT^{-1}$
with some coefficients $a,b$, while for $T<T_c$,
$F$ is constant \cite{derrida1981random}. 
This $F$ is a monotonic function of $T$ as expected.
It would be interesting to see if the same prescription works in the present case
of random matrix model.
We leave this as an interesting future problem.

\section{Discussion}\label{sec:discussion}
In this paper we have analyzed the quenched free energy in Gaussian matrix model
directly without using the replica method.
We find an integral representation of the exact quenched free energy \eqref{eq:F-int}.
The exact quenched free energy is a monotonic function of temperature as expected,
and the entropy computed from this free energy approaches $\log N$ at high temperature
and vanishes at zero temperature.

There are many interesting open questions. It is very interesting to
see if there is a phase transition in the large $N$ limit.
In the case of random energy model, 
it is known that there is a phase transition
associated with the replica symmetry breaking and the low temperature
phase corresponds to a spin glass \cite{gross1984simplest}.
Since the random matrix model considered in this paper
can be thought of as a generalization of the random energy
model, it is tempting to speculate that the quenched free energy 
of the random matrix model
also exhibits a phase transition.\footnote{As discussed in \cite{Okuyama:2019xvg},
all contributions in the decomposition \eqref{eq:Zin-Zconn} become comparable
around $T\sim N^{-2/3}$ and the dominance of disconnected part $\bra Z(\bt)\ket^n$
is lost below this temperature.
In the strict $N\to\infty$ limit this temperature $T\sim N^{-2/3}$ vanishes.
To keep this scale finite, we can take a scaling limit $N\to\infty,T\to0$ with
$TN^{2/3}$ fixed. As discussed in \cite{Engelhardt:2020qpv}, this amounts to focusing on the edge of the Wigner distribution \eqref{eq:wigner} and this scaling limit corresponds to the so-called Airy limit.} 
To settle this issue it is important to understand the analytic continuation
of $\bra Z(\bt)^n\ket$ to $n<1$. We proposed a simple condition \eqref{eq:condition}
for the analytic continuation of $\bra Z(\bt)^n\ket$ to satisfy.

It would be very interesting to generalize our analysis to the JT gravity matrix model
and see if the spin glass phase is realized at low temperature \cite{Engelhardt:2020qpv}.
In \cite{Engelhardt:2020qpv} the quenched free energy is computed 
by a certain prescription
of the analytic continuation
of $\bra Z(\bt)^n\ket$ and it leads to a pathological behavior at low temperature.
It is argued in \cite{Johnson:2020mwi} that this problem is resolved
by including the non-perturbative effect.
It would be very interesting to complete the program of replica computation
of the free energy in JT gravity.

Our analysis suggests that at low temperature the smallest eigenvalue
(or the lowest energy state) gives a dominant contribution to the quenched free energy.
This reminds us of the ``eigenbrane'' introduced in \cite{Blommaert:2019wfy}.
Perhaps the spacetime picture of the low temperature
phase is described by an eigenbrane with one of the eigenvalues pinned to the 
edge of the spectral density.
It would be interesting to investigate this picture further.

\acknowledgments
This work was supported in part by JSPS KAKENHI Grant No. 19K03845.

\end{document}